\documentclass[square]{ws-procs11x85}

\usepackage{multicol,float} 

\begin{document}

\title{On the X-ray/TeV connection in Galactic jet sources}

\author{V. Bosch-Ramon}
\address{Max Planck Institut f\"ur Kernphysik\\
Saupfercheckweg 1, Heidelberg 69117, Germany\\
E-mail: vbosch@mpi-hd.mpg.de}

\author{D. Khangulyan}
\address{Max Planck Institut f\"ur Kernphysik\\
Saupfercheckweg 1, Heidelberg 69117, Germany\\
E-mail: Dmitry.Khangulyan@mpi-hd.mpg.de}

\author{F.~A. Aharonian}
\address{Max Planck Institut f\"ur Kernphysik\\
Saupfercheckweg 1, Heidelberg 69117, Germany\\
E-mail: Felix.Aharonian@mpi-hd.mpg.de; 
Dublin Institute for Advanced Studies, Dublin, Ireland}

\begin{abstract} 
There are three Galactic jet sources, from which TeV emission has been detected:
LS~5039, LS~I~+61~303 and Cygnus~X-1. These three sources show power-law tails at X-rays and soft gamma-rays
that could indicate a non-thermal origin for this radiation. In addition, all three sources apparently show
correlated and complex behavior at X-ray and TeV energies. In some cases, this complex behavior is 
related to the orbital motion (e.g.
LS~5039, LS~I~+61~303), and in some others it is 
related to some transient event occurring in the system (e.g.
Cygnus~X-1, and likely also LS~I~+61~303 and LS~5039). Based on modeling or energetic grounds, it seems
difficult to explain the emission in the X-/soft gamma-ray and the TeV bands as coming from the same
region (i.e. one-zone). We also point out the importance of the pair creation phenomena in these systems,
which harbor a massive and hot star, for the radio and the X-ray emission, since a secondary pair
radiation component may be significant in these energy ranges. 
Finally, we discuss that in fact the presence of the star can indeed
have strong impact on, beside the non-thermal radiation production, the jet dynamics.
\end{abstract}


\bodymatter

\begin{multicols}{2}

\section{Introduction}\label{aba:sec1}

Three X-ray binary systems presenting extended radio emission have been detected in the TeV range: LS~5039 \cite{aharonian05}; LS~I~+61~303
\cite{albert06}; and Cygnus~X-1 \cite{albert07}. These three sources seem to show similarities between the X-ray and the TeV lightcurves
(LS~5039: see fig.~3 and fig.~5 in [\refcite{bosch05}] and [\refcite{aharonian06}], respectively; LS~I~+61~303: see fig.~3 in
[\refcite{chernyakova06}]; Cygnus~X-1: see fig.~4 in [\refcite{albert07}]). In addition, LS~5039 show apparently similar photon index/flux
changes in both energy ranges \cite{aharonian06}. In the case of LS~5039 and LS~I~+61~303, the radiation variability seems to be associated
with the orbital motion \cite{aharonian06,albert06}. Otherwise, Cygnus~X-1 was detected during a transient event, and a possible orbital 
link cannot be neither stated nor discarded; short TeV flares have been also detected in LS~5039 and LS~I~+61~303 (see
[\refcite{nauroistalk}] and [\refcite{ricostalk}], respectively). The short ($\sim$~hours) TeV flares observed in these three sources could
have X-ray counterparts, since active X-ray states (quasi) simultaneous with TeV activity have been reported: possibly in LS~5039
\cite{horns06}, in LS~I~+61~303 \cite{esposito07,paredestalk}, and in Cygnus~X-1 \cite{albert07}. 

\section{On the X-ray/TeV connection and the importance of the primary star}\label{aba:sec1}

\begin{figure*}[] 
\centerline{\psfig{file=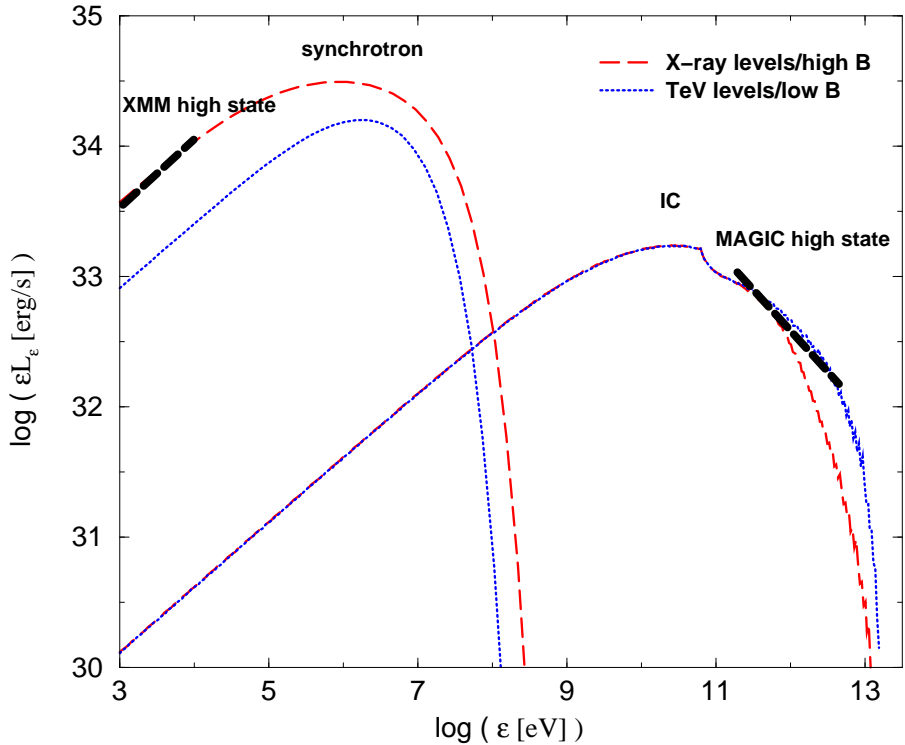,width=12cm}} \caption{Computed spectral energy distribution of 
the synchrotron and IC emission produced in LS~I~+61~303. We show two cases the only difference 
between them is the magnetic field ($B$) value: a case in which the X-ray 
emission is explained in the context of synchrotron radiation ($B=2.3$~G; dotted line); a case in which 
the TeV emission is explained in the context of IC radiation ($B=0.9$~G long-dashed line). The luminosity 
injected in relativistic particles is $2\times 10^{36}$~erg~s$^{-1}$, and the acceleration efficiency is 
$\sim 0.1\,qBc$, enough to explain the observed maximum photon energy. For a description of the used one-zone 
model and the system parameters, see [\refcite{bosch06}]. From the curves, it is seen that to explain the highest energy
band of the very high-energy spectrum \cite{albert06} requires X-ray fluxes below the observed ones \cite{sidoli06}.} \label{lsi}
\end{figure*}

\subsection{The X-ray and the TeV emission}

The link between the X-ray and the TeV emission in all these three sources is certainly far from being clear.
Despite showing similar behavior, the region producing the radiation in these two energy bands is
likely different, as explained in this section.

In the case of LS~5039, strong theoretical arguments (basically: the expected low magnetic field in the emitter; see
[\refcite{khangulyan07}]) show that the X-ray and the TeV emission cannot come from the same electron
population\footnote{Even if the TeV radiation has a hadronic origin, a leptonic population different from the secondary
pairs produced in hadronic interactions would be required to explain the X-ray fluxes.}, although some sort of 
physical link would
be required to explain the similar lightcurve and photon index evolution along the orbit. 

Regarding LS~I~+61~303, it is also difficult to reconcile its X-ray and TeV radiation during the phases with a
one-zone model if we want to explain the highest energy band of the TeV spectrum at the same time as the
hard X-ray fluxes (see Fig.~\ref{lsi}). 

Finally, concerning Cygnus~X-1, the emission in the TeV range cannot be produced by the same electrons that emit hard
X-rays because the emitting processes are expected to be different; the hard X-rays would come from the base of the jet
or from a corona-like region (see, e.g. [\refcite{markoff05,maccarone05}]); the TeV could be produced either by hadronic
processes or via inverse Compton (IC) where non-thermal particles interact with the targets likely provided by the star
(i.e. wind ions \cite{romero03} or UV photons \cite{bednarek07}). Soft gamma-ray radiation showing a power-law-like
steep spectrum has been found in Cygnus~X-1 \cite{mcconnell02}. Unfortunately, there are no soft gamma-ray data
simultaneous with the TeV observations, preventing from a meaningful comparison. Nevertheless, we note that if 
this soft gamma-ray emission were
produced in the jet, an injection luminosity of relativistic electrons equivalent to the kinetic 
luminosity of the jet \cite{gallo05}
would be needed, unless the injected electron distribution had a extremely high low-energy cutoff ($\sim$~TeV).

In short, although there should be a physical link between the X-ray and the TeV emission in LS~5039 and LS~I~+61~303, the emitter modeling
must go beyond the one-zone approximation. Even in the case of Cygnus~X-1, to explain the soft gamma-rays and the TeV emission requires a
very peculiar particle distribution with an unrealistic low-energy cutoff of $\sim$~TeV, incompatible with TeV observations \cite{albert07}.
Otherwise, the required relativistic particle luminosity is hardly tenable.

\subsection{The role of the star}

\begin{figure*}[] 
\centerline{\psfig{file=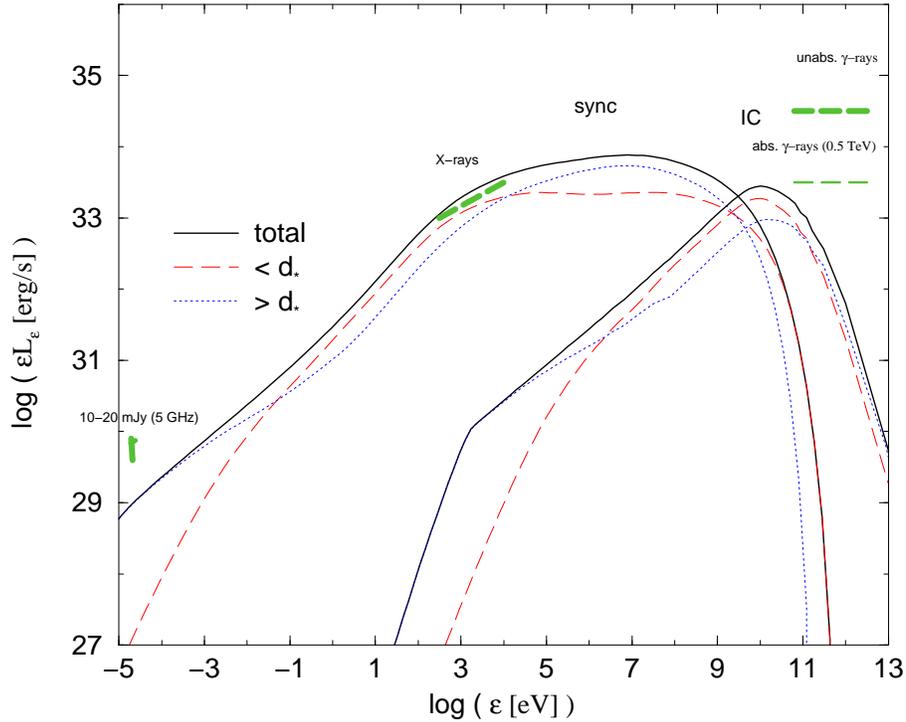,width=12cm}} \caption{Computed spectral energy distribution of 
the synchrotron and IC emission produced by the secondary pairs created in the photon field of the primary 
OB star. The adopted parameters correspond to an object with characteristics similar to those of 
LS~5039 and Cygnus~X-1 \cite{bosch07b}. Concerning (isotropic) injected TeV luminosity, we take 
$3\times 10^{35}$~erg~s$^{-1}$. The magnetic field in the system is taken to be $B=100\,(R_*/R)$~G, where 
$R_*$ is the stellar radius. The advection speed at which particles are transported by the wind 
is $2\times 10^8$~cm~s$^{-1}$. We note that the radio fluxes are not far from those observed in microquasars, 
showing that the secondary radio emission may be significant. 
Also, the X-ray flux is similar to that observed 
in LS~5039 \cite{bosch07}.} \label{sec}
\end{figure*}

To understand the X-ray/TeV connection in these systems, there are additional complexities that are to be taken into
account before looking for a physical explanation. The presence of the massive hot star, with its strong photon and
magnetic field, and strong stellar wind, could have a significant impact in the emission and absorption processes, and
on the physics of the jet or outflow, which is commonly assumed to be the accelerator/emitter in the sources discussed here. For
instance, in Fig.~\ref{sec} it is shown the emission generated by electron-positron  pairs produced via pair creation
in the photon field of the primary star under reasonable assumptions for the system environment. Such a secondary
radiation could be a significant fraction, if not a dominant one, of the non-thermal radiation produced in TeV close
binary systems \cite{bosch07b}. In addition, the stellar wind ram pressure could distort the jet dynamics significantly,
producing also non-thermal emission via the generation of shocks \cite{peruchoposter}.

\section{Discussion}

The aim of this work is to remark the complexity of the phenomena that can lead to the production of non-thermal emission in microquasars,
and galactic compact sources in general, which harbor a massive hot star. This complexity arises either based: on modeling grounds (i.e.
observations cannot be easily explained); or on an exploration of the importance of several elements, which are unavoidably playing some role
in the considered scenario (i.e. the stellar radiation field and wind). 

In some cases the TeV radiation and its evolution could be explained via geometrical effects related to the emission and absorption processes
(e.g. LS~5039\cite{khangulyan07}). However, the similar X-ray and TeV behaviors could be additionally pointing to underlying variability of
the non-thermal emitter intrinsic properties (e.g. acceleration efficiency, magnetic field, relativistic particle energy budget, etc.). In
addition, the spectra and fluxes of the radiation in the X-ray and the TeV band seem to be incompatible with a simple approach considering
only one electron population (or even one hadronic-secondary leptonic population). Finally, the fact that the emitter is embedded in a
powerful material outflow coming from the primary star likely implies that {\it isolated} jet models can hardly work to explain the
radio-to-TeV emission.  This could already happen at the first order approximation, or even at the zero one. All this should be taken into
account when trying to generalize the behavior of these sources, generalization that seems far from trivial.

\section{Summary}

The X-ray and the TeV emission from the three jet Galactic sources detected up to now at very high energies shows a very complex behavior and
cannot be explained in the simple context of one-zone models. The presence of the primary star cannot be neglected regarding the non-thermal
processes occurring in (and the dynamics of) the jet and its surroundings.

\section*{Acknowledgments}
V.B-R. gratefully acknowledges support from the Alexander von Humboldt Foundation. V.B-R. acknowledges support
by DGI of MEC under grant AYA2007-68034-C03-01, as well as partial support by the European Regional
Development Fund (ERDF/FEDER).

\end{multicols}


\begin{thebibliography}{9}

\bibitem{aharonian05} Aharonian, F.A. {\it et al.}, {\em Science}  {\bf 309},
746 (2005).

\bibitem{albert06}
Albert, J. {\it et al.}, {\em Science}, {\bf 312}, 1771 (2006).

\bibitem{albert07}
Albert, J. {\it et al.}, {\em ApJ}, {\bf 665}, 51 (2007).

\bibitem{bosch05}
Bosch-Ramon, V., Paredes, J.~M., Rib\'o, M. {\it et al.}, {\em ApJ}, {\bf 628}, 388  (2005).

\bibitem{aharonian06}
Aharonian, F.~A. {\it et al.}, {\em A\&A}, {\bf 460}, 743 (2006).

\bibitem{chernyakova06}
Chernyakova, M., Neronov, A., \& Walter, R. {\em MNRAS}, {\bf 372}, 1585 (2006).

\bibitem{nauroistalk}
de Naurois, M. for the HESS collaboration, Talk presented in the conference: The keV to TeV connection, Rome,
October 2006.

\bibitem{ricostalk}
Rico, J. for the MAGIC collaboration, this conference.

\bibitem{horns06}
Horns, D. for the HESS collaboration, 2nd Workshop On TeV Particle Astrophysics, Madison, August 2006.

\bibitem{esposito07}
Esposito, P., Caraveo, P.~A., Pellizzoni, A. {\it et al.} {\em A\&A}, {\bf 474}, 575 (2007).

\bibitem{talk}
Paredes, J.~M., this conference [astro-ph/0803.1097]

\bibitem{khangulyan07} 
Khangulyan, D., Aharonian, F.~A., Bosch-Ramon, V. {\em MNRAS} {\bf 383} 467  (2008).

\bibitem{markoff05}
Markoff, S., Nowak, M.~A., Wilms, J. {\em ApJ}, {\em 635}, 1203 (2005).

\bibitem{maccarone05}
Maccarone, T.~J., {\em MNRAS}, {\bf 360}, L68 (2005).

\bibitem{romero03} 
Romero, G. E., Torres, D. F., Kaufman Bernad\'o, M. M., Mirabel, I. F. 
{\em A\&A}, {\bf 410}, L1 (2003).

\bibitem{bednarek07} 
Bednarek, W. \& Giovannelli, F. {\em A\&A}, {\bf 464}, 437 (2007).

\bibitem{mcconnell02}
McConnell, M.~L., Zdziarski, A.~A., Bennett, K. {\em ApJ}, {\bf 572}, 984 (2002).

\bibitem{gallo05}
Gallo, E., Fender, R., Kaiser, C. {\em Nature}, {\bf 436}, 819 (2005).

\bibitem{bosch07b}
Bosch-Ramon, V., Khangulyan, D., Aharonian, F.~A. {\em A\&A}, in press (2008).

\bibitem{peruchoposter}
Perucho, M. \& Bosch-Ramon, this conference [astro-ph/0711.3556].

\bibitem{bosch06} 
Bosch-Ramon, V., Paredes, J.~M., Romero, G.~E., \& Ribó, M. {\em A\&A}, {\bf 459}, L25 (2006).

\bibitem{sidoli06} 
Sidoli, L., Pellizzoni, A., Vercellone, S., {\it et al.}, {\em A\&A}, {\bf 459}, 901 (2006).

\bibitem{bosch07}
Bosch-Ramon, V., Motch, C., \& Rib\'o, M. {\it et al.}, {\em A\&A}, {\bf 473}, 545 (2007). 

\end{thebibliography}
\end{document}